\documentclass[aps,twocolumn,showpacs]{revtex4}
\usepackage{graphicx}
\begin{document}

\title{A one-dimensional Fermi accelerator model with moving wall
described by a nonlinear van der Pol oscillator}

\author{$^1$Tiago Botari and $^{2,3}$Edson D.\ Leonel}

\affiliation{$^1$Departamento de F\'isica - UNESP - Univ Estadual
Paulista, Av.24A, 1515 -- Bela Vista -- CEP: 13506-900 -- Rio Claro -- SP
-- Brazil\\
$^2$Departamento de Estat\'{\i}stica  Matem\'{a}tica
Aplicada e Computa\c c\~{a}o - UNESP - Univ Estadual Paulista\\
Av.24A, 1515 -- Bela Vista -- CEP: 13506-900 -- Rio Claro -- SP --
Brazil\\$^3$Abdus Salam ICTP, 34100 Trieste, Italy}

\date{\today} \widetext


\begin{abstract}
A modification of the one-dimensional Fermi accelerator model is
considered in this work. The dynamics of a classical particle of mass $m$,
confined to bounce elastically between two rigid walls where one is
described by a non-linear van der Pol type oscillator while the other one
is fixed, working as a re-injection mechanism of the particle for a next
collision, is carefully made by the use of a two-dimensional non-linear
mapping. Two cases are considered: (i) the situation where the particle
has mass negligible as compared to the mass of the moving wall and does
not affect the motion of it; (ii) the case where collisions of the
particle does affect the movement of the moving wall. For case (i) the
phase space is of mixed type leading us to observe a scaling of the
average velocity as a function of the parameter ($\chi$) controlling the
non-linearity of the moving wall. For large $\chi$, a diffusion on the
velocity is observed leading us to conclude that Fermi acceleration is
taking place. On the other hand for case (ii), the motion of the moving
wall is affected by collisions with the particle. However due to the
properties of the van der Pol oscillation, the moving wall relaxes again
to a limit cycle. Such kind of motion absorbs part of the energy of the
particle leading to a suppression of the unlimited energy gain as
observed in case (i). The phase space shows a set of attractors of
different periods whose basin of attraction has a complicate organization.
\end{abstract}

\maketitle
\section{Introduction}
\label{sec1}

As firstly proposed by Enrico Fermi \cite{Ref1} as an attempt to describe
the high energy of cosmic particles interacting with moving magnetic
clouds, the Fermi accelerator model consists of a classical particle of
mass $m$ (denoting the cosmic particle) confined to bounce between two
rigid walls. One is periodically moving in time (making allusions to the
moving magnetic clouds) while the other one is fixed (working as a
returning mechanism for a next collision with the moving wall). The phase
space of the model is defined by the type of the motion of the moving
wall. For a smoothly periodic motion, say a sinusoidally function,
periodic islands, chaotic seas and a set of invariant KAM curves
are all observed coexisting in the phase space. The diffusion in the
velocity is limited by the KAM curves and, deceptively, Fermi acceleration
(unlimited energy growth) is not observed. On the other hand when the
motion of the moving wall is considered of saw-tooth type, the mixed
structure of the phase space is not observed anymore and unlimited energy
is observed. The interest in Fermi acceleration has then increased and
several applications have been observed in different areas of science
including astrophysics \cite{Ref2,Ref3}, plasma physics \cite{Ref4}, 
optics \cite{Ref5,Ref6}, atomic physics \cite{Ref7} and even in time
dependent billiard problems \cite{Ref8}. The traditional approach to
describe Fermi acceleration developing in such type of time dependent
system is generally given in terms of a diffusion process which takes
place in momentum space. The evolution of the probability density function
for the magnitude of particle velocities as a function of the number of
collisions is determined by the Fokker-Planck equation and results for
the one-dimensional case consider either static wall approximation and
moving boundary description \cite{Ref9,Ref10,Ref11}. The phenomenon
however seems not to be robust since dissipation is assumed to be a
mechanism to suppress Fermi acceleration \cite{Ref12}.

In this paper we revisit the one-dimensional Fermi accelerator model
however considering the motion of the moving wall given by a van der Pol
type and considering two cases: (i) the mass of the particle is
negligible as compared to mass of the moving wall; (ii) the collisions of
the particle affect the motion of the moving wall, which are restored to
a limit cycle after a relaxation time. Our main goal is to understand and
describe the influences of a limit cycle type motion of the moving wall to
the dynamics of the particle and hence to the properties of the average
velocity in the phase space. To the best knowledge of the authors, this
is the first time this approach is considered in this model and the
results may have applications to different set of models, particularly
to the class of time-dependent billiard problems. The model then consists
of a classical particle of mass $m$ which suffers collisions with two
walls. One is moving in time whose motion is described by a van der Pol
equation leading to a limit cycle dynamics while the other one is fixed
and works as a returning mechanism of the particle to a next collision
with the moving wall. The dynamics is constructed by a two dimensional
nonlinear mapping for the variables velocity of the particle and time.
For case (i) the dynamics leads to a mixed phase space structure where
periodic islands are observed surrounded by chaotic seas and limited by a
set of invariant KAM curves. As soon as the parameter $\chi$ controlling
the non-linearity of the moving wall raises, the position of the lowest
invariant KAM curve raises too leading to an increase in the average
velocity of the particle. Scaling arguments are used to describe the
behavior of the average velocity as a function of the parameter $\chi$
and a set of critical exponents are obtained. On the other hand for case
(ii), the collisions of the particle with the moving wall indeed affect
the dynamics of such wall, leading it out/in of the limit cycle. After a
relaxation time however, the moving wall reaches the limit cycle again. A
set of different periodic attractors is observed in the phase space and
the organization of the basin of attraction of each attractor shows to be
complicate. A histogram of periodic orbits is also constructed leading us
to observe a high incidence of low period orbits as compared to large
period.

This paper is organized as follows. In Sec. \ref{sec2} we construct the
model and describe the equations that give the dynamics of the motion.
Section \ref{sec3} is devoted to discuss the case of negligible mass of
the particle including results of the phase space, characterization of
chaotic orbits and scaling of the average velocity. The case where
collisions of the particle affect the motion of the moving wall is
described in Sec. \ref{sec4}. Conclusions and final remarks are drawn in
Sec. \ref{sec5}.


\section{The model and the mapping}
\label{sec2}

In this section we present all the details needed for the construction of
the mapping. The model consists of a classical particle of mass $m$ which
is confined to bounce between two walls. One of the them is assumed to be
fixed at position $x=L$ and collisions are assumed to be elastic. The
other one moves in time and the oscillations are described by a van der
Pol oscillator whose average position is $x=0$. We describe two situations
where: (i) the mass of the particle is sufficiently small as compared to
the mass of the moving wall and, (ii) the case where the mass of the
particle is not negligible therefore affecting the dynamics of the
moving wall.

\begin{figure}[t]
\centerline{\includegraphics[width=1.0\linewidth]{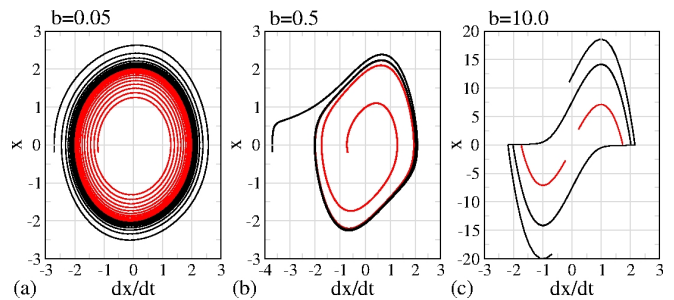}}
\caption{{(Color online) Phase space plot for a van der Pol oscillator
considering the parameters $M=k=x_0=1$ and: (a) $b=0.05$; (b) $b=0.5$ and;
(c) $b=10$.}}
\label{Fig1}
\end{figure}

Indeed, the van der Pol oscillator is a generalization of a harmonic
oscillator that contains a nonlinear and dissipative term.
Applications of the van der Pol oscillator can be found in many different
systems including dusty plasma \cite{Ref13,Ref14}, coupled oscillators
\cite{Ref15,Ref16} and many others. The differential equation describing
the van der Pol oscillator is defined as
\begin{equation}
M \frac{d^2x}{dt^2} + b(x^2-{x_0} ^2)\frac{dx}{dt}+kx =
F_0\sin(\omega_ft)~,
\label{Eq_1}
\end{equation}
where $M$ is the mass, $b$ denotes the dissipative term and $k$ is the
Hook's constant. The nonlinear term $(x^2-{x_0} ^2)$ creates a competition
between adding and subtracting energy to/from the system. When
$|x|<|x_0|$ there is a frictional drag force while for $|x|>|x_0|$
there is a negative friction. This competition has a balance of energy
such that the sum of the added and reduced energy in a cycle is zero,
therefore creating a periodic motion which is commonly called in the
literature as limit cycle. When $b\approx0$ there is a sinusoidal
motion with a quasi-circular limit cycle of radius $2x_0$. With the growth
of $b$, the limit cycle changes to a particular form and the time of
relaxation of the oscillator becomes shorter. Figure \ref{Fig1} shows a
typical phase space obtained from numerical integration of Eq.
(\ref{Eq_1}), by using the Gauss-Radau \cite{Ref17} integrator, for the
parameters $M=k=x_0=1$, $F_0=0$ and: (a) $b=0.05$; (b) $b=0.5$ and; (c)
$b=10$.

To describe the model itself, we have to define a set of dimensionless and
therefore more convenient variables. Defining  $y= x/L$ and
$\tau=t\omega_0$ the equation is rewritten as
\begin{equation}
\frac{d^2y}{d{\tau}^2} + \chi({y}^2-{y_0}^2)\frac{dy}{d\tau}+ y =0.
\label{vanderpol}
\end{equation}
where $\chi = b/(M \omega_0 L)$ and $y_0=x_0/L$ and $F_0$ was set as null.
With this new variables, the distance between the average position of the
moving wall and fixed wall becomes dimensionless and equals $1$. 

In our description and to determine the law that describes the collisions,
we assumed that the momentum and energy are conserved at the instant of
each collision. With this approach, the velocities of the particle and
the moving wall after the $(n+1)^{th}$ collision are 
\begin{eqnarray}
v^p_{n+1} = \frac{\mu -1}{1+\mu} (v^p_{n}-v^w_n) + v^w_n,\nonumber\\
v^w_{n+1} = \frac{2\mu }{1+\mu}(v^p_{n}-v^w_n) + v^w_n.
\label{vpvw}
\end{eqnarray}
where the upper index $p$ and $w$ stand for the particle and moving wall
respectively. Here $\mu=m/M$ denotes the ratio of mass of the particle and
the mass of the moving wall. Therefore the particle and the moving wall
change energy and momentum upon collision. In the general case, after the
collision of the particle with the moving wall and, due to the change of
energy, the velocity of the moving wall changes therefore bringing the
moving wall out of the limit cycle. However and depending on the control
parameters, the limit cycle is approached again asymptotically. Therefore
we have to consider two separate cases: (i) $\mu=0$ and; (ii) $\mu\ne 0$.
For case (i), i.e., when the mass of the particle is sufficiently small as
compared to the mass of the moving wall $m\ll M$ leading to the limit of
$\mu=0$. For this case we assumed that either energy and velocity of the
moving wall are not changed after the collision, therefore
$v^w_{n+1}=v^w_{n}$. For case (ii) where $\mu>0$, the collisions of the
particle with the moving wall indeed affect the motion of the moving
wall bringing it in or out of the limit cycle. It however relaxes again
to the limit cycle as time evolves. We therefore consider the two cases
in separate sections.

\section{The case of $\mu=0$}
\label{sec3}

\begin{figure*}[t]
\begin{center}
\centerline{\includegraphics[width=16cm,height=12cm]{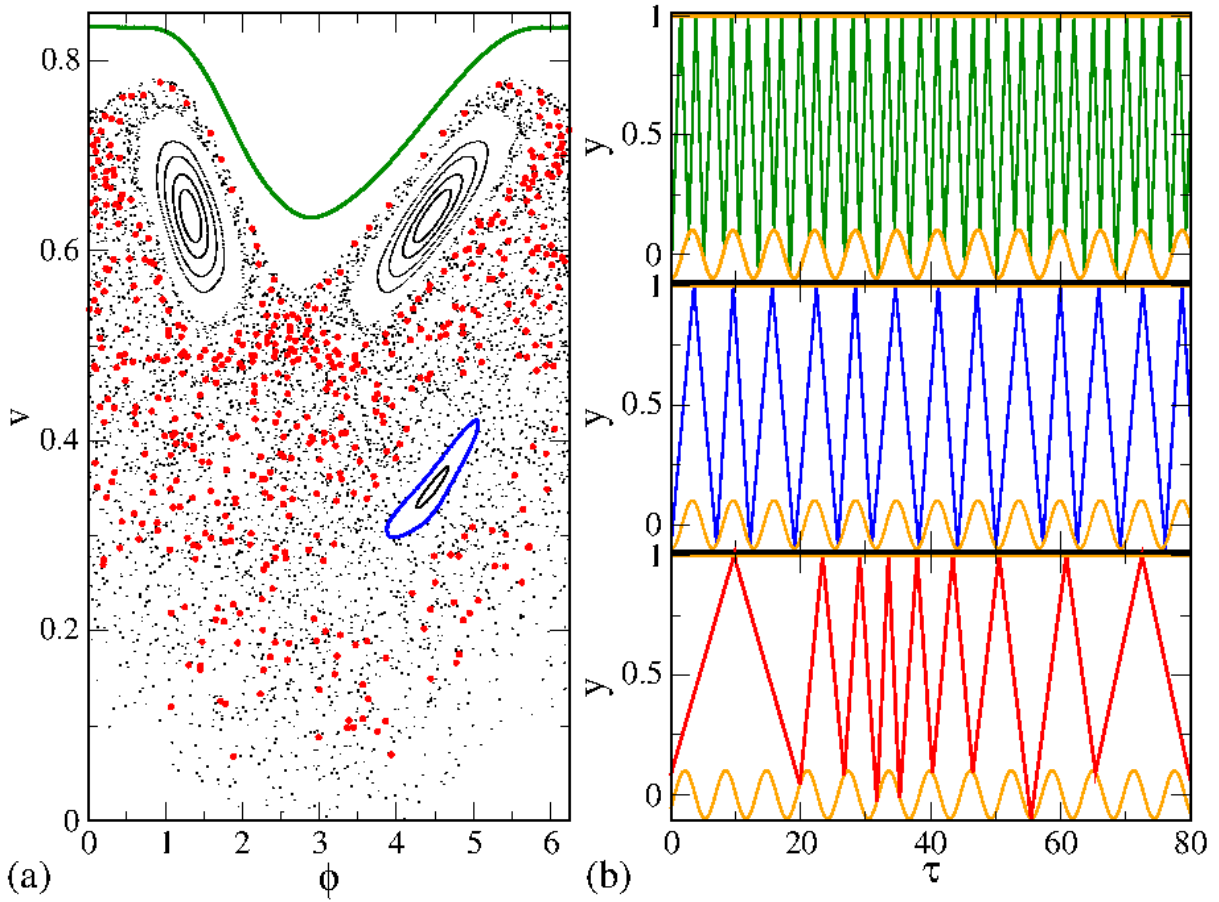}}
\end{center}
\caption{{(Color online) (a) Phase space for mapping (\ref{mapamu=0}) for
the parameters $y_0=0.05$ and $\chi=0$. (b) Show the three types of
behavior observed in (a): green for KAM curves, blue for periodic islands
and red for chaotic dynamics.}}
\label{Fig2}
\end{figure*}

In this section we discuss the case of $\mu=0$. It implies that the
velocity of the moving wall is not affected by the change of energy with
the particle i.e., $v^w_{n+1}=v^w_{n}$. Moreover, once relaxed to the
limit cycle, the moving wall stays in such a regime forever. For such a
dynamical regime, we can also calculate a period of oscillation $T$. The
dynamics is then described by a two-dimensional mapping for the variables
velocity of the particle and phase of the moving wall
\begin{eqnarray}
\phi_{n+1} &=& [\phi_n +\Delta t_{n+1}]~ mod ~ T,\nonumber\\
v_{n+1} &=&  2v_w(\phi_{n+1})\pm v_n,
\label{mapamu=0}
\end{eqnarray}
where the $+$ sign of second term stands for successive collisions while
$-$ denotes non successive collisions. A successive collision is defined
as a collision that the particle has with the moving wall without leaving
the collision zone, i.e., without been reflected backwards by the fixed
wall. Here $\phi_n$ is the phase of the moving wall while $v_n$ is the
velocity of the particle and the index $n$ denotes the instant of the
$n^{th}$ collision with the moving wall. The term $v_w(\phi)$ represents
the velocity of the moving wall that is obtained numerically from the
integration of the van der Pol oscillator (Eq. \ref{vanderpol}). It
indeed has two relevant control parameters namely, $y_0$ and $\chi$.
The parameter $y_0$ controls the amplitude of the limit cycle and $\chi$
controls the amplitude of nonlinear dissipative term. The term $\Delta
t_{n+1}$ is solved numerically from the equation
\begin{eqnarray*}
\Delta t_{n+1} = \frac{2-x_w(\phi_n) - x_w(\phi_{n+1})}{v_n},\\
\end{eqnarray*}
for successive collisions and
\begin{eqnarray*}
\Delta t_{n+1} = \frac{x_w(\phi_{n+1}) - x_w(\phi_n)}{v_n},
\end{eqnarray*}
for non successive collisions. For the case of $\mu=0$, the mapping
preserves the following measure in the phase space $du={(v-v_w(\phi))d\phi
dv}$.

The phase space for the case of $\mu=0$ is mixed and periodic islands,
KAM curves and a chaotic sea are observed, as shown in Fig. \ref{Fig2}
(a). The colors in the three plots of Fig. \ref{Fig2}(b), showing
trajectory of the particle as a function of time, denote the three types
of behavior observed in (a): (i) green for KAM curves; (ii) blue for
periodic islands and; (iii) red for chaotic dynamics. Indeed the dynamics
along a KAM curve is quasi-regular leading to a small variance of the
velocity of the particle while all phases are in principle allowed
leading to a seemingly closed curve as shown in Fig. \ref{Fig2}(a). For
periodic islands, i.e., the case of Fig. \ref{Fig2}(b) the orbit has a
regular motion and visits only a finite portion of the space including a
specific region in phase. The last case (iii) of chaotic motion shows an
apparently {\it erratic} trajectory in the phase space.

\begin{figure}[t]
\centerline{\includegraphics[width=1.0\linewidth]{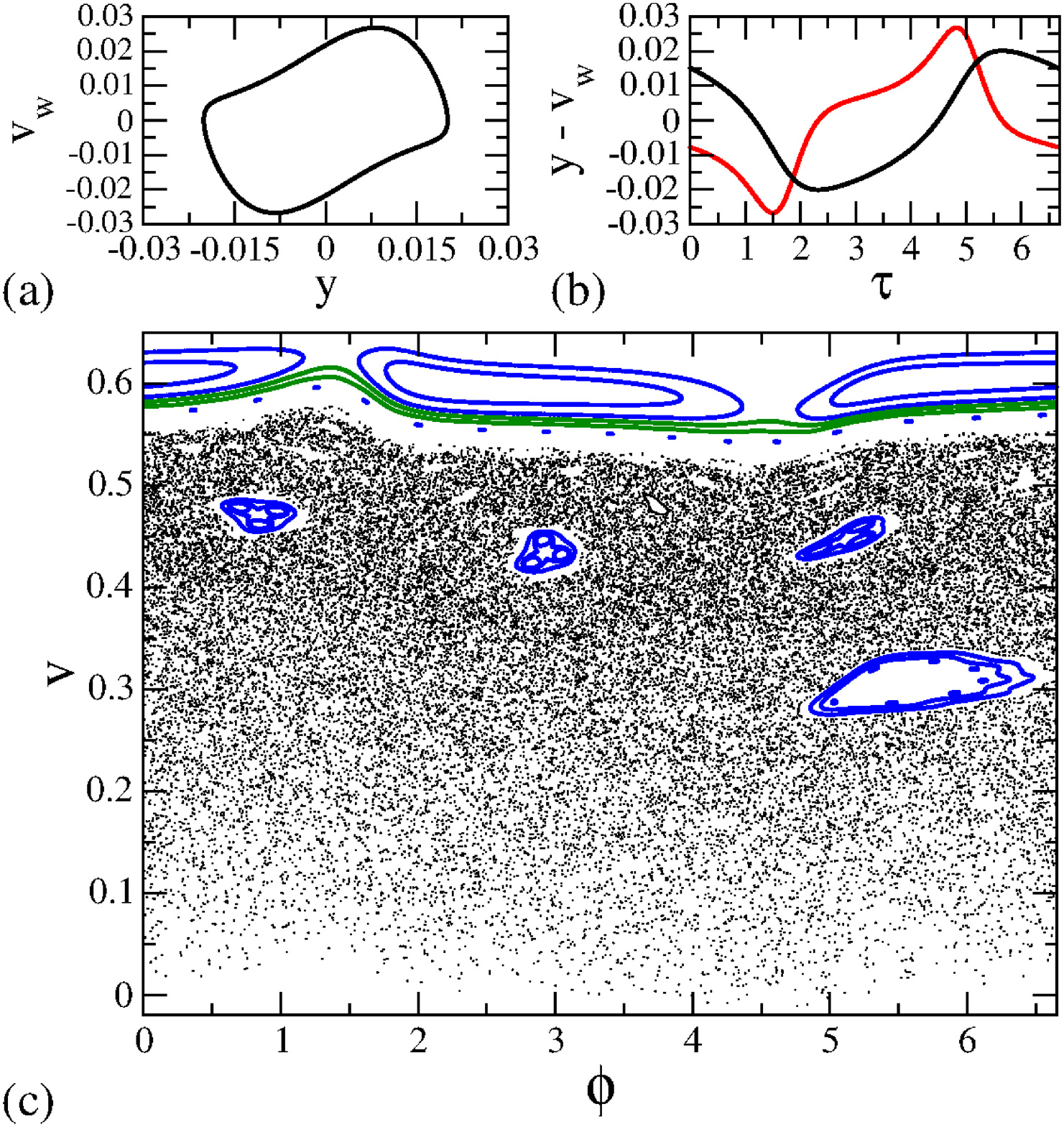}}
\caption{{(Color online) (a) Plot of the phase space for moving wall,
$v_w~versus~y$; (b) plot of $v_w~versus~\tau$ (red line) and
$y~versus~\tau$ (black line); (c) phase space for the mapping
(\ref{mapamu=0}) $v~versus~\phi$. The control parameters used were
$\chi=10^4$ and $y_0=0.01$.}}
\label{Fig3}
\end{figure}

Let us now discuss some properties of the phase space. Indeed the
parameter $\chi$ controls the shape of the limit cycle. For $\chi\approx
0$ the present model recovers the results of the Fermi-Ulam model (FUM)
\cite{Ref18}. In the FUM, the moving wall is described by a sinusoidally
function and, as expected, the phase space is mixed. A plot of the phase
space for the parameters $y_0=0.05$ and $\chi=0$ is shown in Fig.
\ref{Fig2}(a). With the increase of $\chi$, the shape of the limit cycle
changes. Figure \ref{Fig3} shows
the plots of: (a) the phase space for moving wall, $v_w~versus~y$; (b)
plot of $v_w~versus~\tau$ (red line) and $y~versus~\tau$ (black line)
while (c) shows a plot of phase space for the mapping (\ref{mapamu=0})
$v~versus~\phi$. The control parameters used in all plots were $\chi=10^4$
and $y_0=0.01$.

\begin{figure}[t]
\centerline{\includegraphics[width=1.0\linewidth]{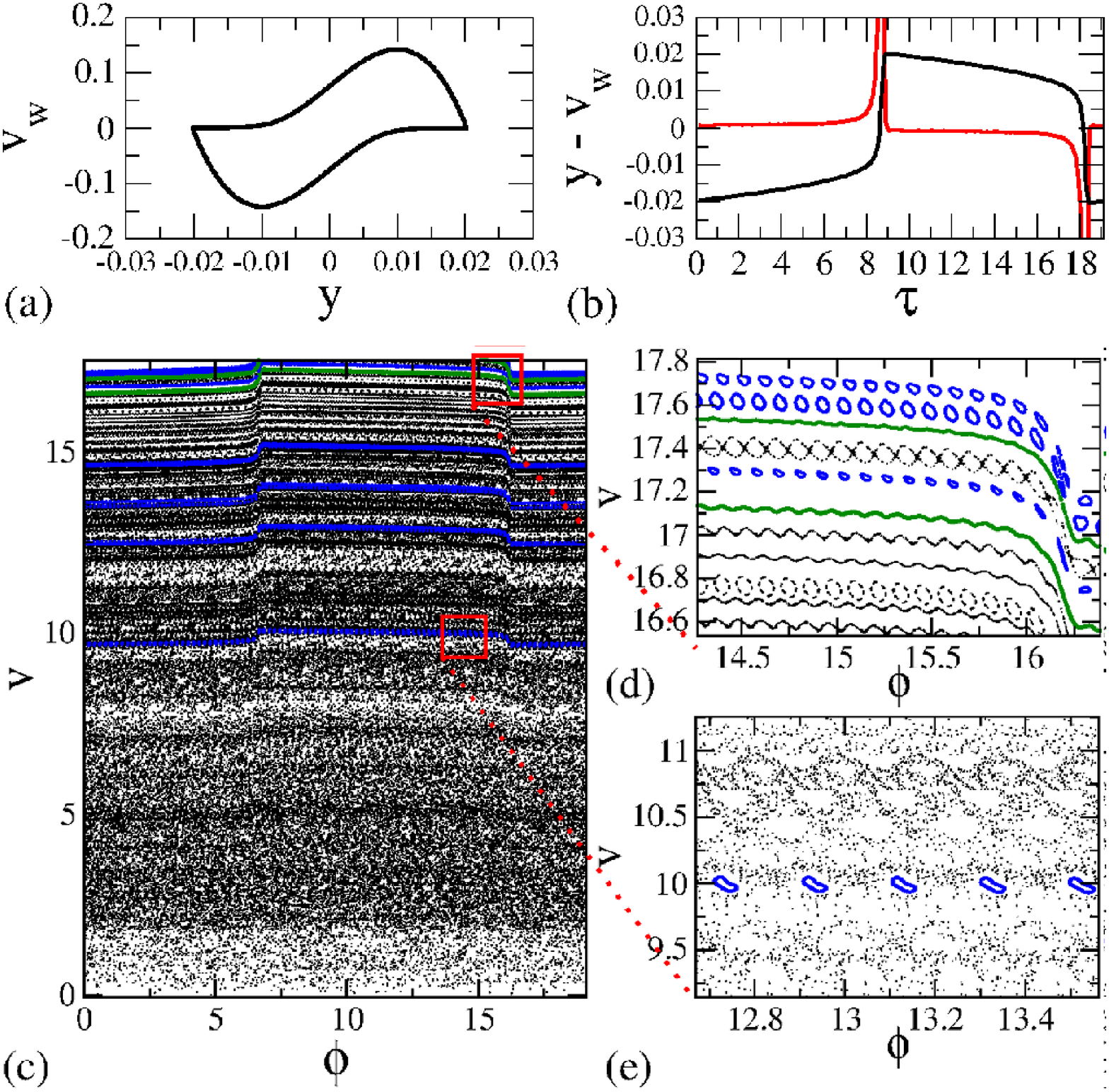}}
\caption{{(Color online) (a) Plot of the phase space for moving wall,
$v_w~versus~y$; (b) plot of $v_w~versus~\tau$ (red line) and
$y~versus~\tau$ (black line); (c) phase space for the mapping
(\ref{mapamu=0}) $v~versus~\phi$.  The control parameters used were
$\chi=10^5$ and $y_0=0.01$.}}
\label{Fig4}
\end{figure}

For a large value of the parameter $\chi$, for example $\chi=10^5$, the
phase space generated from mapping (\ref{mapamu=0}) has, at first sight, a
strange form where a not conventional structure is clearer shown. Figure
\ref{Fig4} 
shows the corresponding plots of: (a) the phase space for moving wall,
$v_w~versus~y$; (b) plot of $v_w~versus~\tau$ (red line) and
$y~versus~\tau$ (black line) while (c) shows a plot of phase space for the
mapping (\ref{mapamu=0}) $v~versus~\phi$. Indeed a zoom-in shown in (d)
and (e) exhibit in a minor scale, the structures expected to be observed
in mixed phase space.
In the reference \cite{video1} a video is shown to
demonstrate the change of parameter $\chi$ cause in phase space of
the particle.
 
The modifications caused in the dynamics of the particle due to the
increase of the parameter $\chi$ may affect some observables in the phase
space. Particularly the position of the lowest KAM curve is raised by
a raise in $\chi$, leading the particle to acquires more energy from the
moving wall. In what follows, our investigation is as function of the
parameter $\chi$. To do so, we chose to describe the behavior of the
average velocity of the particle along the chaotic orbits as well as the
behavior of the positive Lyapunov exponent.

Let us start with the average velocity. To do so, we consider the
evolution of the average velocity of the particle as a function of the
number of collisions $n$ as well as the parameters $y_0$ and $\chi$. The
average velocity is obtained as
\begin{eqnarray}
<v(n)>=\frac{1}{K}\sum_{i=1}^{K} v_i(n),
\label{rugnormal}
\end{eqnarray}
where $K$ denotes the total number of initial conditions (ensemble of
different particles) and $v_i(n)$ is the average velocity obtained along
the orbit for a single initial condition. It is defined as
\begin{eqnarray}
v_i(n)=\frac{1}{n}\sum_{j=1}^{n} v(j)
\end{eqnarray}
where the index $i$ corresponds to one orbit of the ensemble $K$. This
kind of ensemble average is widely used in many different dynamical
systems \cite{Ref18,Ref19,Ref20,Ref21,Ref22}. Several tests were made and
shown that a results with low fluctuation is obtained with an ensemble of
initial condition $K$ of the order $500$ to $1000$. The simulations
involved in the calculation of $<v(n)>$ are extremely time consuming
especially due to the numerical solution of Eq. \ref{vanderpol}. As an
attempt to speed up the simulations, we introduced a new method to
calculate the average velocity. Indeed it can be used because the mapping
preserves a measure in the phase space and, due to Poincar\'e recurrence
theorem, the particle may returns to close to a neighboring region to
where it started as one wants.

\begin{figure}[t]
\centerline{\includegraphics[width=1.0\linewidth]{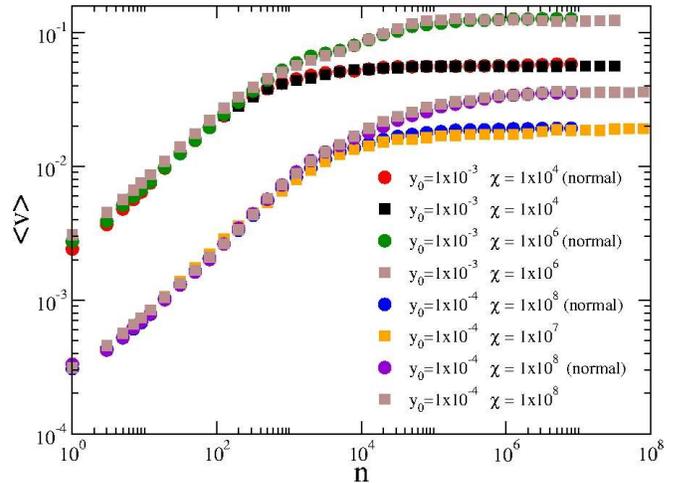}}
\caption{(Color online) Plot of $<v>~versus~n$ for different control
parameters and initial conditions, as labeled in the figure. Bullets
denote the traditional (normal) method of evolving an ensemble of initial
conditions and squared correspond to the proposed method.}
\label{figCom}
\end{figure}

Suppose a time series $v(n)$ is given, where $n=1,2...N$. We define $v_m$
as a sup value of velocity for region of low energy in the phase space
that belongs to the chaotic sea. Indeed $v_m$ should be larger than the
velocity of the moving wall and at the same time should be smaller as
compared to first KAM curve. Then we construct a vector $k(i)$ where $k$
is a series from each $n$ for those velocities where the condition
$v(n)<v_m$ is observed such that $i$ is an index that represents the times
this will occur. Thus we define the transformations
\begin{equation}
v_i(n)=\frac{1}{n}\sum_{j=1}^{n-1} v(j-k(i)),
\end{equation}
and
\begin{equation}
<v>(n)=\lim_{N\rightarrow\infty}\frac{1}{K'}\sum_{i=1}^{K'} v_i(n),
\end{equation}
where $K'=dim(k)$, and $dim$ denotes the dimension of vector $k$. Using
this procedure, the evolution of a single orbit for a long time is enough
to obtain the average properties and therefore to speed up the simulations
as compared to the normal method of evolving a long ensemble of different
initial conditions. To have an idea of the accuracy of the proposed
method, Fig. \ref{figCom}
shows a plot of $<v>~versus~n$ for different control parameters and
initial conditions, as labeled in the figure. Bullets correspond to
the traditional (normal) method of evolving an ensemble of initial
conditions \cite{Ref18} while squares denote the proposed method. We see
that curve generated by the two methods remarkably overlap each other
confirming the procedure can be used. Of course some cautions were taken
as for example when the particle is trapped in a sticky region. In this
case the problem is fixed by just increasing the number of collisions or
by a change the initial condition.

\begin{figure}[t]
\centerline{\includegraphics[width=1.0\linewidth]{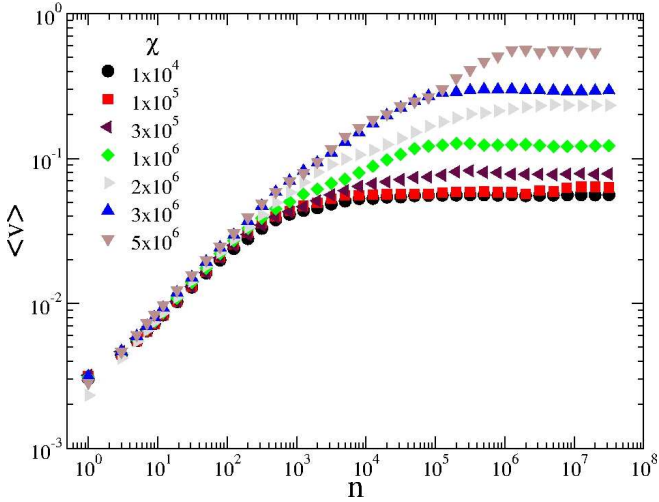}}
\caption{(Color online) Plot of the average velocity as a
function of $n$ for the parameter $y_0=0.001$ and different values of
$\chi$, as labeled in the figure.}
\label{fig5}
\end{figure}

\begin{figure}[t]
\centerline{\includegraphics[width=1.0\linewidth]{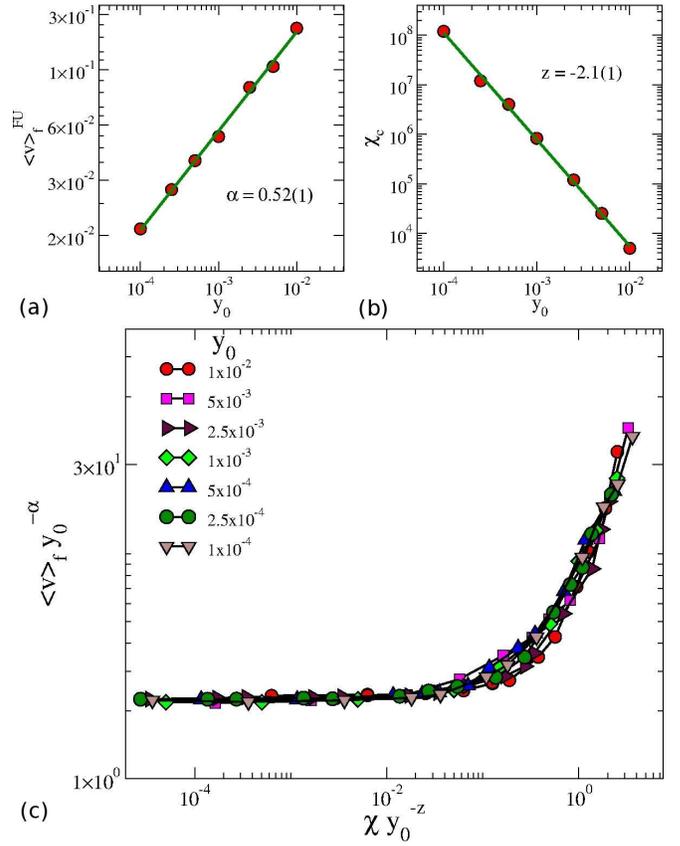}}
\caption{{(Color online) (a) Plot of $<v>_f^{FU}~versus~y_0$. A
fitting furnishes an exponent $\alpha=0.52(2)$; (b) Plot of
$\chi_c~versus~y_0$. A numerical fit gives an exponent $z=-2.1(2)$; (c)
Overlap of different curves of $<v>$ onto a single plot, after a suitable
rescale of the axis, for different control parameters, as labeled in the
figure.}}
\label{fig6}
\end{figure}

\begin{figure}[t]
\centerline{\includegraphics[width=1.0\linewidth]{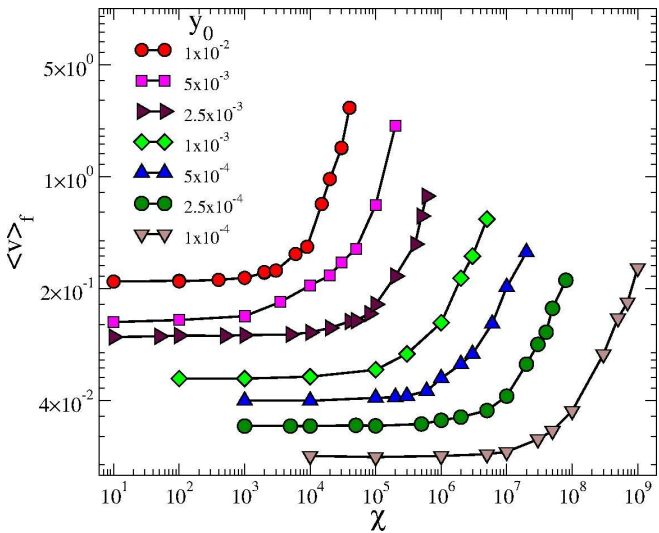}}
\caption{{(Color online) Plot of the final average velocity as a
function of $\chi$ for different values of $y_0$, as labeled in the
figure.}}
\label{Fig6new}
\end{figure}

Established the procedure to obtain the average velocity, let us now
discuss some properties as a function of the control parameters. Indeed,
starting with an initial condition with low velocity, the average velocity
$<v>$ grows for small values of $n$ and, after passing by a crossover
regime, it becomes constant for large enough $n$. The changeover from
growth to the regime marked by a constant plateau defines a final average
velocity $<v>_f$. Figure \ref{fig5}
shows different plots of $<v>~versus~n$ for the parameter $y_0=0.001$ and
different values of $\chi$, as labeled in the figure. For $\chi\approx 0$,
the results obtained recover those already known for the traditional
FUM \cite{Ref18}. As shown in Fig. \ref{fig5}, a change in $\chi$ leads
the asymptotic dynamics for large $n$ to reach different saturation of the
velocity. It is expected that a similar behavior is to be observed as the
parameter $y_0$ varies and it indeed happens. A plot of the final average
velocity $<v>_f$ as a function of the parameter $y_0$ is shown in Fig.
\ref{fig6}(a).
The behavior of the final average velocity as a function of $\chi$ is
shown in Fig. \ref{Fig6new}
for different values of $y_0$, as labeled in the figure. We see from this
figure that the final average velocity stays in a plateau for a large
range of $\chi$ and that it depends on $y_0$. After a critical
parameter $\chi_c$ is reached, the average velocity starts to increase
with a power law. Based on the behavior observed in Fig. \ref{Fig6new},
we propose the following:
\begin{enumerate}
\item{For $\chi \ll {\chi_c}$, the average velocity $<v>_f$ behaves as
\begin{equation}
<v>_f(\chi) \propto y_0^{\alpha}~,
\label{Eq5}
\end{equation}
where $\alpha$ is a critical exponent;}
\item{For $\chi \gg {\chi_c}$, the average velocity is written as
\begin{equation}
<v>_f(\chi) \propto \chi^{\beta}~,
\label{Eq6}
\end{equation}
where $\beta$ is also a critical exponent;}
\item{Finally, the crossover $\chi_c$ that marks the change from
the plateau to the regime of growth is given by
\begin{equation}
\chi_c \propto y_0^{z}~,
\label{Eq7}
\end{equation}
where $z$ is called a dynamic exponent.}
\end{enumerate}

After considering these three initial suppositions, the description of
the asymptotic average velocity $<v>_f$ may be made in terms of a scaling
function of the type
\begin{equation}
<v>_f (\chi ,y_0)=l<v>_f(l^a \chi , l^b y_0)~,
\label{Eq8}
\end{equation}
where $l$ is the scaling factor, $a$ and $b$ are scaling exponents. If
we chose properly the scaling factor $l$, it is possible to relate the
scaling exponents $a$ and $b$ with the critical exponents $\alpha$,
$\beta$ and $z$. Choosing $l^a\chi =1$, which leads $l=\chi^{-1/a}$, Eq.
(\ref{Eq8}) is rewritten as
\begin{equation}
<v>_f (\chi ,y_0)=\chi^{-1/a} <v>_f(1, \chi^{-b/a}y_0)~.
\label{Eq9}
\end{equation}
Comparing Eqs. (\ref{Eq6}) and (\ref{Eq9}), we obtain $\beta=-1/a$.

Choosing now $l^by_0=1$, we have $l=y_0^{-1/b}$ and Eq. (\ref{Eq8}) is
written as
\begin{equation}
<v>_f (\chi ,y_0)=y_0^{-1/b}<v>_f(y_0^{-a/b}\chi,1).
\label{Eq10}
\end{equation}
An immediate comparison of Eqs. (\ref{Eq5}) and (\ref{Eq10}) gives us
$\alpha=-1/b$. Given the two different expressions of the scaling
factor $l$, it is easy to obtain a relation for the dynamic exponent
$z$, that is given by
\begin{equation}
z={\alpha\over\beta}~.
\label{Eq15}
\end{equation}
The critical exponents $\alpha$, $\beta$  and $z$ can be obtained via
extensive numerical simulations. For the initial regime of plateau for
$\chi \ll \chi_c$, a power law fitting to the plot of $<v>_f^{p}~versus~
y_0$ gives that $\alpha=0.52(1)$. On the other hand, the regime of growth
for the curves of $<v>_f~versus~\chi$, i.e., for $\chi \gg \chi_c$, we
obtain that $\beta=1.2(3)$  Finally, a fitting to the plot of
$\chi_c~versus~y_0$ gives that $z=-2.1(1)$. Figure \ref{fig6}(a,b) shows
the behavior of $<v>_f^{FU}~versus~y_0$ and $\chi_c~versus~y_0$
respectively.

Given that the critical exponents are now obtained, a suitable rescale of
the axis from the plot shown in Fig. \ref{Fig6new} can be made to overlap
all curves of $<v>_f$ onto a single and seemingly universal plot, as
shown in Fig. \ref{fig6}(c). This overlap confirms the behavior of the
average velocity at the asymptotic dynamics is scaling invariant with
respect to both $y_0$ and $\chi$.

Given we have described the behavior of the average velocity, let us now
concentrate to discuss the Lyapunov exponent for chaotic orbits. Indeed
the Lyapunov exponent gives the exponential rate of divergence or
convergence of the evolution of close initial conditions in phase space.
Thus when at least one Lyapunov exponent $\lambda$ positive, the system
has a chaotic component. The Lyapunov exponent is defined as \cite{Ref23}
\begin{equation}
\lambda_i = \lim_{n\rightarrow \infty} \frac{1}{n} \ln|\Lambda_i|
\end{equation}
where $\Lambda_i$ is the eigenvalues of matrix $M=\prod_{j=1}^n J_j$ where
$J_j$ is the Jacobian matrix evaluated along the orbit $(v_j,\phi_j)$.

\begin{figure}[t]
\centerline{\includegraphics[width=1.0\linewidth]{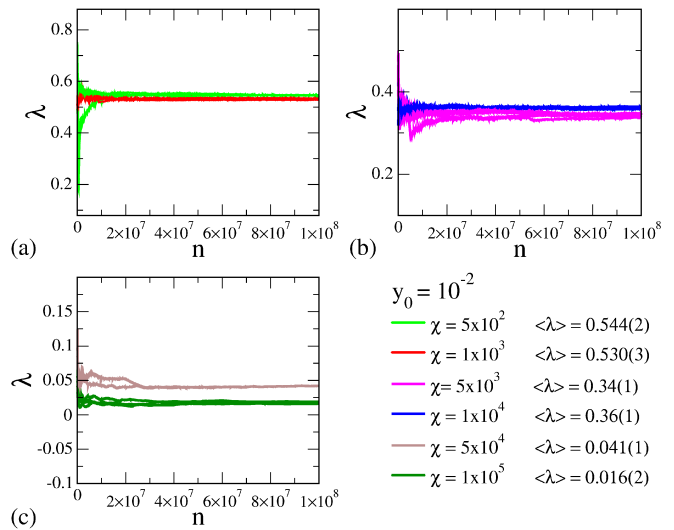}}
\caption{{(Color online) Plot of the Lyapunov exponent averaged
over the chaotic sea for mapping (\ref{mapamu=0}) considering the control
parameter $y_0=0.01$ and: (a)  $\chi = 5\times10^2$ and 
$\chi = 1\times10^3$  ; (b) $\chi = 5\times10^3$ and $\chi = 1\times10^4$;
(c) $\chi = 5\times10^4$ and $\chi = 1\times10^5$.}}
\label{Fig9}
\end{figure}

For the model described by the mapping \ref{mapamu=0} and considering a
fixed value of $y_0$, the positive Lyapunov exponent averaged along the
chaotic sea in the regime of low energy decreases as the control parameter
$\chi$ increases. Figure \ref{Fig9} 
shows the behavior of the Lyapunov exponent as a function of the
collisions with the moving wall for the parameter $y_0=0.01$ and: (a)
$\chi = 5\times10^2$ and $\chi = 1\times10^3$; (b) $\chi = 5\times10^3$
and $\chi = 1\times10^4$; (c) $\chi = 5\times10^4$ and $\chi =
1\times10^5$.

\begin{figure}[t]
\centerline{\includegraphics[width=1.0\linewidth]{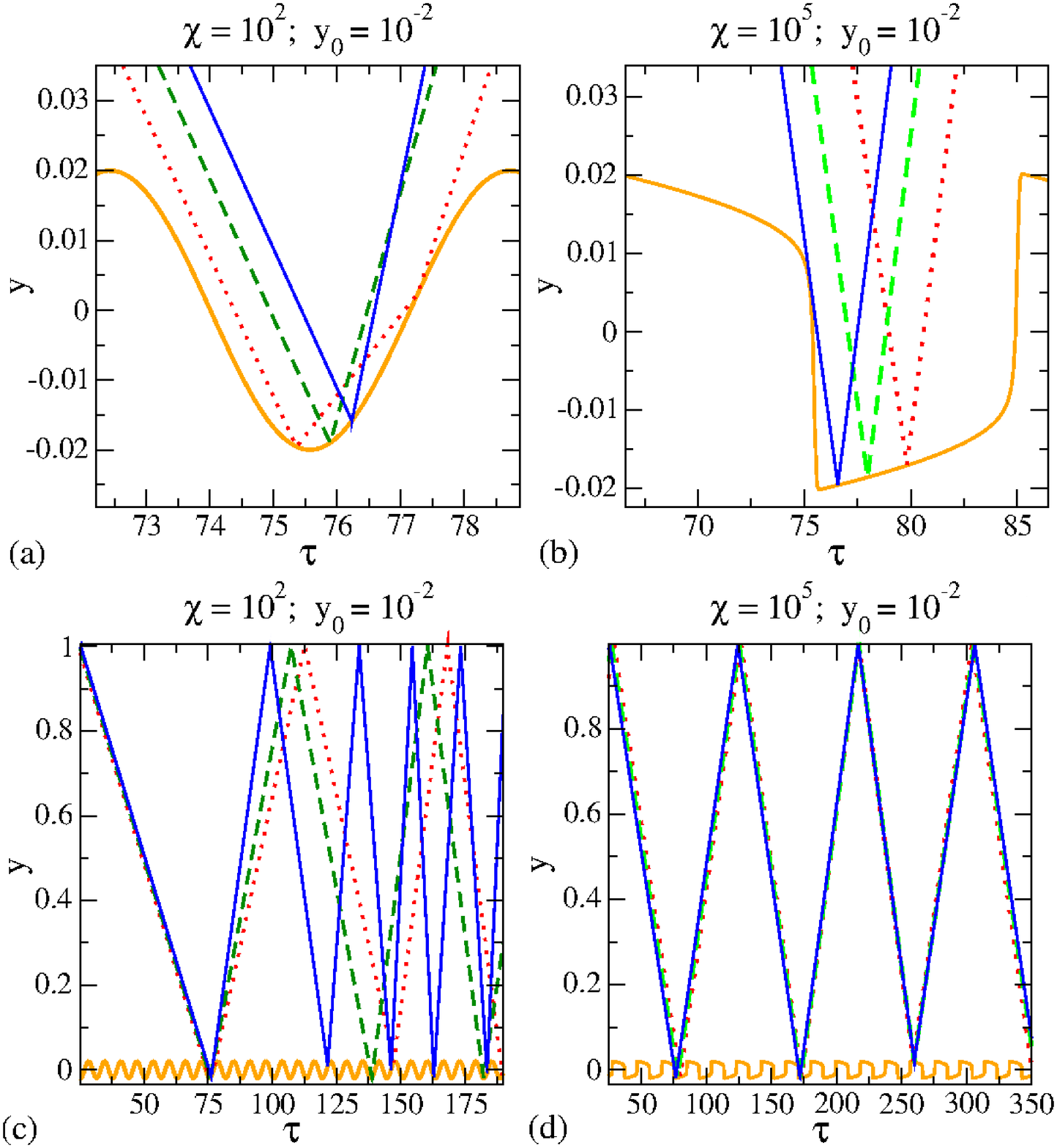}}
\caption{{(Color online) Plot of the trajectory of the particle
as a function of time. The parameters used were $y_0=10^{-2}$ and:
(a) and (c) $\chi =10^2$ and; (b) and (d) $\chi=10^5$.}}
\label{Fig5new}
\end{figure}

A possible explanation of the decrease of the Lyapunov exponent as an
increase of the $\chi$ is due to the shape of the limit cycle. Indeed, for
sufficiently small $\chi$, the shape of form of the limit cycle looks more
like a sine function. However, as the parameter $\chi$ increases, the
shape describing the motion of the moving wall has parts of large
regularity, those where it moves slowly, and parts where it moves very
fast. Therefore, we expected that, as the parameter $\chi$ increases, the
particle suffers more collisions with the {\it regular} motion of moving
wall. Eventually it collides with a region where the moving wall is moving
very fast, therefore leading to a large exchange of energy, producing the
regimes of growing velocities, as discussed before. Figure \ref{Fig5new}
shows plots of the trajectory of the particle as a function of time. The
parameters used were $y_0=10^{-2}$ and: (a) and (c) $\chi =10^2$ and; (b)
and (d) $\chi=10^5$. One can see clearer that in (a) and (c), the
separation of the two neighboring particles is more visible while
compared to (b) and (d).

\section{The case of $\mu\neq0$}
\label{sec4}
Let us consider in this section the case of $\mu\neq0$. When the particle
collides with the moving wall, it perturbs the motion of the moving wall,
bringing it out/in the limit cycle. As the time evolves, the oscillator
pushes the dynamics back to the limit cycle, restoring the dynamics.
Under this circumstance, the mapping is now written as
\begin{eqnarray}
v^p_{n+1} &=& \frac{\mu -1}{1+\mu} (v^p_{n}-v^w_n) + v^w_n,\nonumber\\
v^w_{n+1} &=& \frac{2\mu }{1+\mu}(v^p_{n}-v^w_n) + v^w_n,\nonumber\\
t_{n+1} &=&t_n +\Delta t_{n+1}\nonumber,\\
y_{n+1} &=& y_w(\phi_{n+1})\nonumber.
\label{mapamu}
\end{eqnarray} 
where $v^p$ is the velocity of the particle, $v^w$ is the velocity 
of the moving wall, $\mu$ is the ratio of mass of the particle and
mass of the wall, $t$ is the time, $y$ the position of the particle and
$y^w$ is the position of the moving wall upon collision. The term $\Delta
t_{n+1}$ is obtained as in the same way as made for the case of $\mu=0$
and is obtained numerically upon to an accuracy of $10^{-12}$.

\begin{figure}[t]
\centerline{\includegraphics[width=1.0\linewidth]{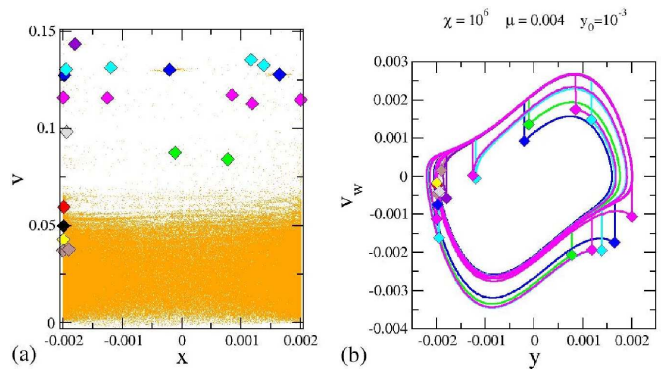}}
\caption{{(Color online) Plot of the phase space $v~versus~x$.
Square represents different attractors, yellow denotes the initial
transient; (b) Phase space for the van der Pol oscillator when it is
perturbed by collisions with the particle. Squares denote the instant of
collision. The parameters used are $y_0=10^{-3}$, $\mu=0.004$ and
$\chi=10^6$.}}
\label{fig11}
\end{figure}

The phase space is now defined by $(y,v,v_w)$, i.e. the position of
moving wall at the instant of the collision $y$, the velocity of the
particle $v$ and the velocity of the moving wall $v_w$. Depending on the
initial conditions and on the set of control parameters, the dynamics
evolves to different fixed points, after passing by a transient.
Therefore, we investigate the basin of attraction for the fixed points
considering the parameters $\chi=10^6$, $y_0=0.001$ and $\mu=0.004$. The
elliptic fixed points observed in the case $\mu=0$, turn into sinks
after a bifurcation and initial conditions close enough converge to them
asymptotically, a shown in Fig. \ref{fig11}. 

\begin{figure}[t]
\centerline{\includegraphics[width=1.0\linewidth]{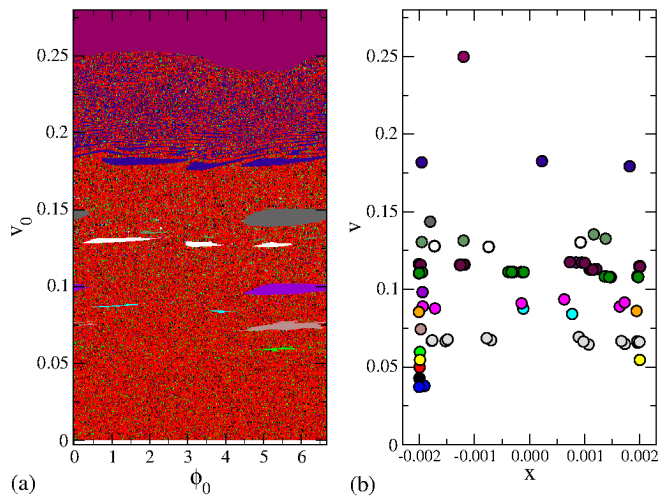}}
\caption{{(Color online) (a) Basin of attraction 
$v_0~versus~\phi_0$; (b) Fixed points $v~verus~x$. The parameters
used are $y_0=10^{-3}$, $\mu=0.004$ and $\chi=10^6$}}
\label{fig12}
\end{figure}

To construct a basin of attraction we set the initial conditions with the
position of the moving wall along the limit cycle which is obtained from
the numerical solution of the integration of the van der Pol oscillator.
Thus the initial condition evolves in time until reaches a regime of
convergence to different attractors. After analyzing the plots of the
basin of attraction, we can identify several attractors with different
periods as shown in Fig. \ref{fig12}.

\begin{figure}[t]
\centerline{\includegraphics[width=1.0\linewidth]{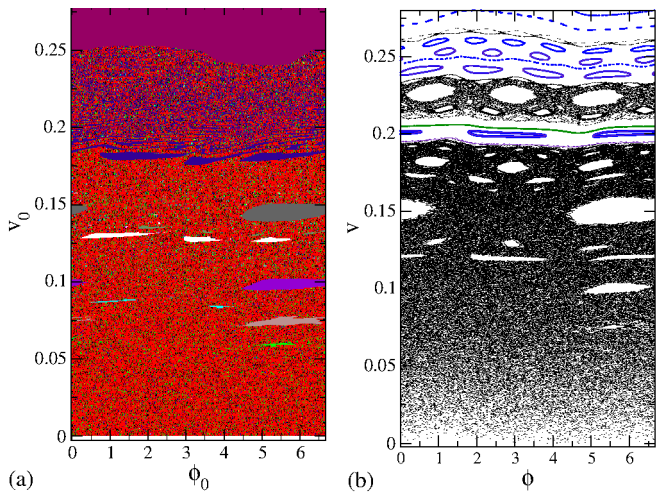}}
\caption{{(Color online) (a) Basin of attraction 
$v_0~versus~\phi_0$ for the parameters $y_0=10^{-3}$, $\mu=0.004$ and
$\chi=10^6$; (b) Phase space plot $v~verus~\phi$ for the parameters
$y_0=10^{-3}$, $\mu=0.004$ and $\chi=0$.}}
\label{fig13}
\end{figure}

Considering $v_0>0.25$ we found only attractors of period $1$. We can
compare the basin of attraction with the phase space in case $\mu=0$ and
note a series of similarities as for example, regions of periodic islands 
became regions that lead the dynamics to a periodic attractor as shown in
Fig. \ref{fig13}.
The parameters used to construct Fig. \ref{fig13} were: (a) $y_0=10^{-3}$,
$\mu=0.004$ and $\chi=10^6$ and; (b) $y_0=10^{-3}$, $\mu=0.004$ and
$\chi=0$. In the reference \cite{video2} a video is shown to
demonstrate the dynamics of the system, compared the phase 
space of particle and the phase space of the moving wall.

\begin{figure}[t]
\vspace*{0.5cm}
\centerline{\includegraphics[width=1.0\linewidth]{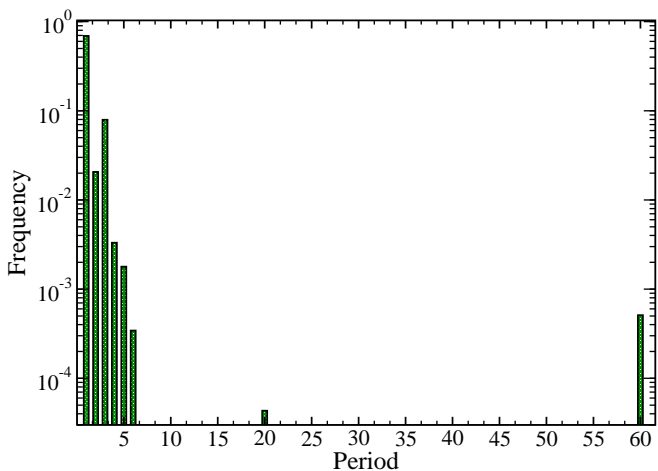}}
\caption{{(Color online) Plot of the histogram of frequency
of initial conditions for the period of the attractor. The parameters
used are $y_0=10^{-3}$, $\mu=0.004$ and $\chi=10^6$.}}
\label{fig14}
\end{figure}

We see from Figs. \ref{fig12} and \ref{fig13} that the basin of
attraction for periodic orbits exhibit quite complicate organization.
To have an estimate of how many initial conditions evolved to period 1,
or period 2 and so on, we constructed a histogram of frequency of
initial conditions that has as final state, a periodic attractor of
period one, or two etc. We found that about $70\%$ of all initial
conditions considered converge to the period 1 sink. The other periods
from $2, 3, 4, 5, 6$ stay around $1\%$ and $10\%$ and some initial
conditions lead to observe periods $20$ and $60$ as shown in Fig
\ref{fig14}. 
The control parameters used in the figure were: $y_0=10^{-3}$, $\mu=0.004$
and $\chi=10^6$.

\section{Conclusion}
\label{sec5}
We revisited and described the dynamics of a classical particle suffering
elastic collisions with two walls. One is fixed and the other one is
moving according to the solution of a van der Pool oscillator. The mapping
describing the dynamics of the model was constructed considering the
cases of: (i) the particle has negligible mass as compared to the mass of
the moving wall and; (ii) the collisions of the particle affect the
dynamics of the moving wall. Due to the properties of the van der Pol
oscillator, after the collisions, the moving wall relaxes again to the
limit cycle. We proposed an alternative method to calculate the average
velocity of the particle along the phase space and, using it, we
investigated the behavior of the average velocity of the particle. Scaling
arguments were used to describe the behavior of the average velocity and
critical exponents $\alpha=0.52(1)$, $\beta=1.2(3)$ and $z=-2.1(1)$ were
used to overlap all curves of average velocity onto a single and seemingly
universal plot. Lyapunov exponents were also discussed as a function of
the control parameters.
 
For case (ii) in which the mass of the particle is taken into account
along the collisions with the moving wall, the dynamics of the system
changes and the system becomes {\it dissipative} leading to appearance of
attractors in the dynamics. The basins of attraction of some attractors
were obtained.
 
As a perspective of the study of this model with such type of
perturbations, a synchronization of the oscillator may be of interest and
applications may be used to control chaos. Other aspect of the model that
can be studied is related to the introduction of two interacting particles
in the system which exchange information through the perturbation caused
by them in the moving wall.


\begin{section}{ACKNOWLEDGMENTS}
TB thanks CAPES and FAPESP. EDL kindly acknowledges the financial support
from CNPq, FAPESP and FUNDUNESP, Brazilian agencies. EDL thanks the
hospitality of the ICTP during his visit. T.B. also thanks Vinicius
Santana, Roberto Eugenio Lagos Monaco and Tadashi Yokoyama for fruitful
discussions. This research was supported by resources supplied by the
Center for Scientific Computing (NCC/GridUNESP) of the S\~ao Paulo State
University (UNESP).
\end{section}

\end{document}